\documentclass[
superscriptaddress,
 aps,
 prx,
floatfix,
reprint,
 amsmath,amssymb,
]{revtex4-2}

\usepackage{graphicx}
\usepackage{dcolumn}
\usepackage{bm}
\usepackage{braket}
\usepackage{siunitx}
\usepackage{hyperref}

\begin{document}

\title{Fast single-atom preparation in optical tweezers via Rydberg blockade}

\author{Yiyi Li}%
\email{These authors contributed equally to this work.}
\affiliation{Department of Electrical and Computer Engineering, Princeton University, Princeton, New Jersey 08544, USA}
\author{Vernon M. Hughes}%
\email{These authors contributed equally to this work.}
\affiliation{Department of Electrical and Computer Engineering, Princeton University, Princeton, New Jersey 08544, USA}
\affiliation{Department of Physics, Princeton University, Princeton, NJ 08544, USA}
\author{Michael Peper}%
\affiliation{Department of Electrical and Computer Engineering, Princeton University, Princeton, New Jersey 08544, USA}
\author{Yicheng Bao}%
\affiliation{Department of Electrical and Computer Engineering, Princeton University, Princeton, New Jersey 08544, USA}
\author{Chenyuan Li}%
\affiliation{Department of Electrical and Computer Engineering, Princeton University, Princeton, New Jersey 08544, USA}
\affiliation{Department of Physics, Princeton University, Princeton, NJ 08544, USA}
\author{Sanzhar Bissenali}%
\affiliation{Department of Electrical and Computer Engineering, Princeton University, Princeton, New Jersey 08544, USA}
\affiliation{California Institute of Technology, Pasadena, California 91125, USA}
\author{Jeff D. Thompson}
\email{jdthompson@princeton.edu}
\affiliation{Department of Electrical and Computer Engineering, Princeton University, Princeton, New Jersey 08544, USA}

\date{\today}

\begin{abstract}
Continuously replenished optical tweezer arrays will unlock unlimited-depth quantum circuits with neutral atom qubits. A key bottleneck limiting the cycle time of these systems is removing atoms from tweezers initially loaded with more than one atom. In the conventional technique of light-assisted collisions, slow collisional dynamics limit the timescale for removing excess atoms to several milliseconds. Here, we propose and demonstrate a scheme for selectively removing one atom at a time from multiply occupied tweezers on a microsecond timescale, using intra-tweezer Rydberg blockade and autoionization. We demonstrate the protocol in $^{171}$Yb in two complementary regimes. 
With two-photon Rydberg excitation from the ground state, we reduce the multi-atom probability to $1\%$ in $\SI{64.8}{\mu s}$, while retaining single atoms in $58.2(2)\%$ of the tweezers, which is comparable to the filling fraction achieved with light-assisted collisions under the same experimental conditions, but over two orders of magnitude faster.
With single-photon excitation from the metastable state $^{3}P_0$, reduced single-atom loss enables a higher filling fraction of $74.8(3)\%$, at the cost of additional temporal overhead to prepare the atoms in $^{3}P_0$. The final filling fraction is limited by an unexplained two-body loss mechanism, which, if solved, could enable fast, quasi-deterministic loading.
\end{abstract}

\maketitle

\section{Introduction}
Neutral atoms in optical tweezer arrays are a promising platform for fault-tolerant quantum computing~\cite{bluvstein2024logical, reichardt2024fault, bedalov2024fault, zhang2025leveragingerasureerrorslogical, mathiot2026benchmarking}, quantum simulation~\cite{labuhn2016tunable,bernien2017probing,browaeys2020many,ebadi2021, Scholl2021quantum, choi2023preparing}, and quantum metrology~\cite{cao2024multi, shaw2024multi}. In all of these applications, quantum operations begin with a deterministically filled array of singly occupied optical tweezers. Preparing such arrays typically involves several stages, including atom loading from a magneto-optical trap~(MOT), removal of multiply occupied tweezers via light-assisted collisions (LAC)~\cite{schlosser2001sub,Schlosser2002Collisional}, measuring the occupied sites, and atom rearrangement~\cite{Barredo2016Anatombyatom, endres2016, kim2016situ}. Historically, atom loading had the longest duration of all of these operations, which created little incentive to make the other steps faster. However, recent advances in fast atomic reloading from a cold reservoir have shortened atom loading times to the millisecond-scale~\cite{Pause2023Reservoir,norcia2024iterative,gyger2024continuous,muniz2025,li2025fast,Chiu2025}, which makes the other steps rate-limiting.

There has been considerable progress on improving the speed of measurement through non-destructive measurements in free-space~\cite{lis2023a,falconi2025microsecond} and cavity arrays~\cite{shaw2026cavity}, as well as destructive measurements~\cite{ma2023a,scholl2023a}. Similarly, many works have studied accelerated rearrangement protocols~\cite{lin2025ai, wang2023accelerating, tian_parallel_2023} and novel optical systems architectures for fast rearrangement~\cite{wei202610megahertz, bytyqi2026device}. However, there has been less progress on improving the speed of single-atom preparation.

Loading atoms into an optical tweezer from a MOT or an atomic reservoir results in a Poisson distribution of atom occupancies. The prevailing approach to preparing singly occupied tweezers is light-assisted collisions~\cite{schlosser2001sub,Schlosser2002Collisional}, which eject atom pairs until the occupancy of each tweezer is either zero or one atom, reaching a steady-state filling fraction of $50-60\%$. As LAC relies on molecular potentials acting over length scales of tens of nanometers, the resulting collisional rates of atoms in optical tweezers are small~\cite{Fuhrmanek2012Lightassisted,pampel2025quantifying,Grun2026Light}, leading to preparation times around \SI{10}{ms} in typical optical tweezers, across many atomic species~\cite{norcia2024iterative,li2025fast,Chiu2025}. By tailoring the optical potentials used for LAC, the collision dynamics can be modified to preferentially remove one atom at a time, increasing the final occupancy at the expense of increasing the preparation time to tens or hundreds of milliseconds~\cite{Grunzweig2010near,Lester2015Rapid,Brown2019Gray,Jenkins2022Ytterbium,zhu2025highefficiencyloading2400ytterbium}. We note that millisecond-scale LAC has been achieved in more tightly confining optical traps including 3D optical lattices~\cite{DePue1999Unity} and nanofiber traps~\cite{Vetsch2012Nanofiber}.

A conceptually distinct strategy exploits the Rydberg blockade for single atom preparation~\cite{Saffman2002Creating,Ebert2014Atomic}. In this scheme, the Rydberg blockade is used to transfer at most one atom into a different hyperfine state, while the remaining atoms are pushed out of the trap. As the Rydberg interaction acts over micrometer length scales, the protocol duration is not limited by the rate of collisions between ground state atoms, and is instead limited by the time required to eject atoms from the trap, which can be as short as 10 microseconds~\cite{bluvstein2024logical}. Using this \emph{keep-one} approach, a single-atom preparation fidelity of \SI{62}{\percent} was achieved~\cite{Ebert2014Atomic}. Extensions to higher single-atom occupancy with larger initial fillings have been proposed, but the practical performance is limited by collisions with nearby ground state atoms, Rydberg \textit{spaghetti} effects, and long-range finite blockade effects~\cite{Ebert2015Coherence, Gaj2014molecular, Derevianko2015Effects}.

In this work, we demonstrate a distinct method of using the Rydberg blockade to prepare singly occupied tweezers, using repeated cycles of Rydberg excitation and autoionization to eject extra atoms one at a time. This \emph{remove-one} approach is fast ---  each cycle of Rydberg excitation and autoionization occurs on the microsecond timescale --- and can reach high loading probabilities by avoiding multi-atom loss in a single cycle. In a $^{171}$Yb tweezer array using two-photon Rydberg excitation from the ground state, we achieve 1\% multi-atom occupancy with 18 cycles of excitation and ejection with a total duration of $\SI{64.8}{\mu s}$, over two orders of magnitude faster than LAC under the same experimental conditions. The single-atom filling fraction is \SI{58.2(2)}{\percent}, which is limited by single-atom loss (\SI{2}{\percent} per cycle) caused by ground-Rydberg decoherence from intermediate state scattering and finite Rydberg lifetime during the Rydberg excitation step. To circumvent this limitation, we apply the same approach using single-photon excitation to a Rydberg state with higher principal quantum number from the metastable state $^3P_0$. This reduces the single-atom loss to 0.26\% per cycle, and increases the final single-atom filling fraction to $\SI{74.8(3)}{\percent}$. However, in this regime, the single-atom occupancy becomes limited by an unexplained two-atom loss process, and the protocol duration extends to several milliseconds, because of the need to pre-cool atoms before transferring to $^3P_0$. This technique requires no additional hardware or control beyond what is already needed for high-fidelity two-qubit gates, and can be implemented in any atomic species where fast Rydberg state removal can be engineered. Future work reducing the unexplained two-atom loss could enable fast, near-deterministic tweezer loading.

\begin{figure}[tb]
\includegraphics[width=\columnwidth]{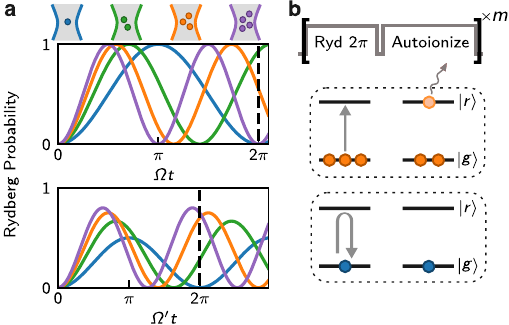}
\caption{\label{fig:fig1} (a)~Collective Rabi oscillations for various atom numbers $N$ (colors indicated in the legend), with on-resonance excitation (top) and detuned excitation (bottom, $\Delta = \Omega$ where the effective Rabi frequency is $\Omega'=\sqrt{\Omega^2+\Delta^2}$). Dashed lines denote single-atom $2\pi$ pulse durations. 
(b)~Scheme for single-atom preparation, involving $m$ cycles with two steps per cycle. First, a Rydberg $2\pi$ pulse excites one atom from a multiply occupied tweezer to the Rydberg state (orange), while keeping the atom in a singly occupied tweezer in the ground state (blue). Second, autoionization removes Rydberg population from the trap.}
\end{figure}

\section{Single-atom preparation using the Rydberg blockade}
The basic mechanism underlying our protocol is illustrated in Fig.~\ref{fig:fig1}. For $N$ atoms in a tweezer within the Rydberg blockade radius~(Fig.~\ref{fig:fig1}(a)), the Rabi frequency of collective excitation to the singly-excited state $\ket{W}=\frac{1}{\sqrt{N}}\sum_{i=1}^N\ket{g_1\cdots r_i \cdots g_N}$ is $\sqrt{N} \Omega$, where $\Omega$ is the single-atom Rabi frequency~\cite{lukin2002blockade}. Applying a pulse with duration $T=2\pi/\Omega$ returns the atom to the ground state when $N=1$, but leaves one atom in the Rydberg state with probability $P(\ket{W})=\frac{1}{2}\Big(1-\cos(2\pi\sqrt{N})\Big)$ when $N>1$. Using autoionization to eject any Rydberg population at the end of the pulse from the trap~\cite{madjarov2020high, burgers2022, Cao2024Autoionization}, we can implement $N\to N-1$ population transfer, while preserving $N=1$ population~(Fig.~\ref{fig:fig1}(b)). The process can be repeated to reduce the multi-atom occupancy to the desired level $P_{\mathrm{th}}$. An on-resonance $2\pi$ pulse also returns all atoms to the ground state when $\sqrt{N}$ is an integer ($N=4,9,\dots$), which would lead to population accumulation at these atom numbers. This problem can be circumvented by detuning the Rydberg excitation by $\Delta$, changing the Rabi frequency scaling to $\sqrt{N\Omega^2 + \Delta^2}$. 

\begin{figure*}[tb]
\includegraphics{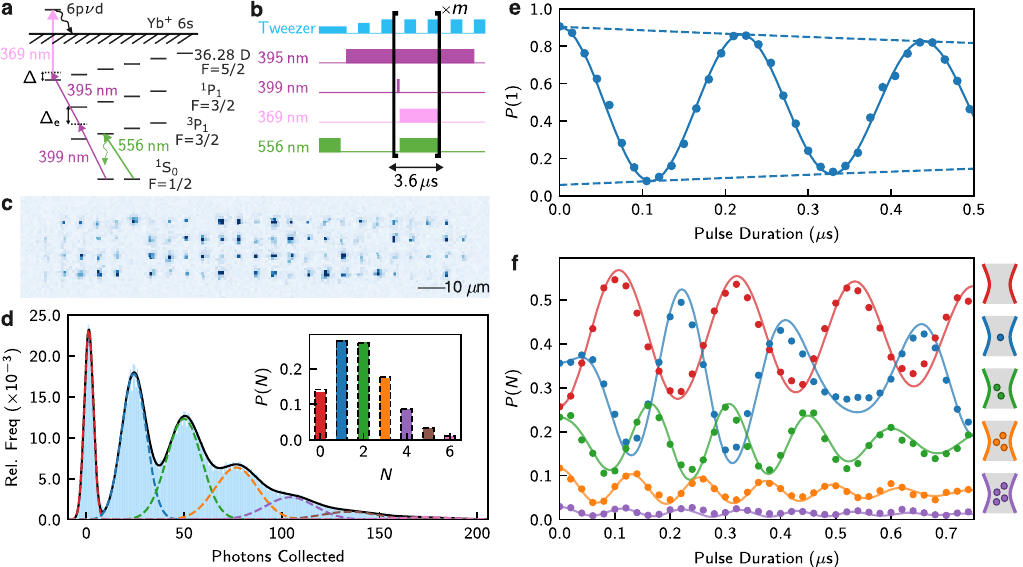}
\caption{\label{fig:fig2} 
(a)~Relevant transitions for the two-photon excitation scheme. The $\SI{556}{nm}$ laser is used for optical pumping, $\SI{399}{nm}$ and $\SI{395}{nm}$ are used for two-photon Rydberg excitations, and $\SI{369}{nm}$ is used for autoionization of the Rydberg state.
(b)~Timing diagram of the protocol for the two-photon excitation scheme. 
(c)~Exemplary single-shot image of the 4$\times$25 tweezer array with number-resolved imaging. 
(d)~Photon-count histogram of number-resolved images recorded after loading  $\overline{N}=1.9$ atoms per tweezer. Dashed lines show Gaussian fits for each atom number $N$, while the solid black line is the cumulative fit. The inset shows the extracted atom-number probability $P(N)$, compared with a Poisson distribution (black dashed bars). 
(e)~Experimental data (circles) and simulation (solid line) of single-atom Rabi oscillations between the ground and Rydberg states, with a Rabi frequency of $\Omega/2\pi=\SI{4.6}{MHz}$ and detuning $\Delta=0$. 
(f)~Multi-atom collective Rabi oscillations. Circles show the measured probability $P(N)$ as a function of Rydberg pulse duration, for an initial average loading of $\overline{N}=1.35$. Solid lines are simulation curves described in the main text. The Rabi frequency is $\Omega/2\pi=\SI{4.7}{MHz}$ and the detuning is $\Delta/2\pi=-\SI{0.4}{MHz}$.
}
\end{figure*}

\begin{figure*}[tb]
\includegraphics{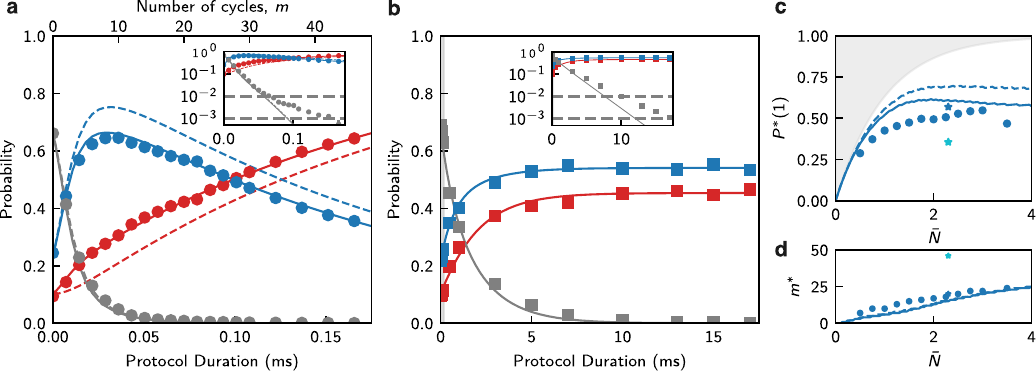}
\caption{
\label{fig:fig3} 
(a)~Experimental results (circles) and simulations (dashed and solid lines) for $P(0)$ (red), $P(1)$ (blue), and $P(>1)$ (gray) as a function of total protocol duration. 
Dashed lines depict master-equation-based simulations described in the main text. Solid lines include an additional $8\%$ probability of $N=2\to0$ loss per pulse.
(b)~Experimental data (squares) and simulations (solid lines) for LAC. The solid lines are based on population evolutions with fitted rates of $\gamma_{1\to 0}=\SI{0(3)}{s^{-1}}$, $\gamma_{2\to 0}=\SI{353(4)}{s^{-1}}$, and $\gamma_{2\to 1}=\SI{177(6)}{s^{-1}}$. The insets in (a) and (b) show the same data on a logarithmic $y$-axis scale.
The average initial number of atoms loaded is $\overline{N}=2.3$ for both (a) and (b). 
(c)~Single-atom filling fraction $P^* (1)$, after reaching $P(>1)<P_{\mathrm{th}}$, as a function of average number of atoms loaded, $\overline{N}$. 
The dark and light blue stars denote $P^*(1)$ for $P_{\mathrm{th}} = 1\%$ and $0.1\%$, respectively, extracted from the data in (a), while the circles depict the experimental results for $P_{\mathrm{th}} = 1\%$ from a different dataset. The dashed (solid) lines depict simulations without (with) the $8\%$ probability of $N=2\to 0$ loss per pulse. The gray shaded region represents the initial $N=0$ probability given by $e^{-\overline{N}}$. 
(d)~The number of cycles required ($m^*$) to reach $P(>1)<P_{\mathrm{th}}$, as a function of average number of atoms loaded, $\overline{N}$. 
}
\end{figure*}

\begin{figure}[t]
\includegraphics{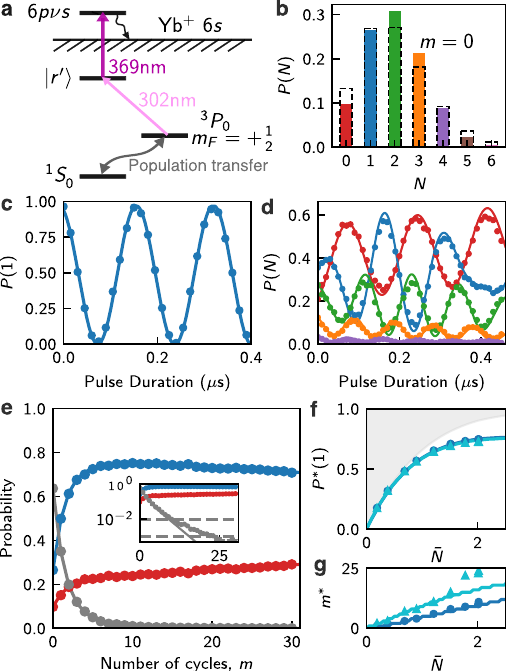} 
\caption{\label{fig:fig4} Single-atom preparation with single-photon Rydberg transitions. (a)~Relevant energy levels for single-photon transitions from the metastable $^3P_0$ state.
(b)~Probability distribution of atom number without applying Rydberg pulses~(colored bars), compared with a Poisson distribution (black dashed bars).
(c)~Exemplary single-atom Rabi oscillations. 
(d)~Multi-atom collective Rabi oscillations, measured with $\overline{N}=1.4$. 
(e)~$P(0)$~(red), $P(1)$~(blue), and $P(>1)$~(gray) as a function of the number of cycles, $m$, starting from $\overline{N}=2.0$. The inset depicts logarithmic scaling of the $y$-axis, with dashed lines showing $P_{\mathrm{th}}=10^{-2}$ and $P_{\mathrm{th}}=10^{-3}$ and solid lines showing the simulation results.
(f)~The single-atom filling fraction, $P^*(1)$, after reaching $P(>1)<P_{\mathrm{th}}$ as a function of the average number of atoms $\overline{N}$ at $m=0$ cycles. The gray shaded region represents the initial $N=0$ probability. 
(g)~The number of cycles required~($m^*$) to reach $P(> 1)<P_{\mathrm{th}}$.
In panels (f) and (g), the dark blue circles denote experimental data with $P_{\mathrm{th}}=1\%$ and the light blue triangles correspond to $P_{\mathrm{th}}=0.1\%$. 
The solid lines in panels (d-g) are based on simulations as described in Appendix~\ref{appendix:3P0}. 
}
\end{figure}

\subsection{Experimental setup}

Using an apparatus previously described in Ref.~\cite{li2025fast}, we first demonstrate the proposed scheme by performing Rydberg excitation from the ground state. In brief, tweezers are loaded from a continuously maintained reservoir of cold $^{171}$Yb atoms and subsequently transported out of the reservoir with a movable tweezer array.
The atoms are then optically pumped into the $\ket{g}=\ket{^1S_0,m_F=-1/2}$ state using 556~nm light on the $^3P_1$ intercombination line~(Fig.~ \ref{fig:fig2}(a)), followed by excitation to the Rydberg state. The Rydberg state $\ket{r} = \ket{36.28,L=2,F=5/2,m_F=-5/2}$ is chosen such that the interaction strength $V$ is large compared to the Rabi frequency for interatomic separations within a tweezer ($V = \SI{275}{MHz}$ at $\sim\SI{1}{\mu m}$), but negligible for atoms in neighboring traps ($V < \SI{7}{kHz}$ at $\SI{6.25}{\mu m}$, see Appendix~\ref{appendix:rydberg_state}). We use a two-photon excitation through an intermediate state $\ket{e}=\ket{6s6p\,^1P_1}$, with counterpropagating beams at wavelengths of $\SI{399}{nm}$ and $\SI{395}{nm}$. Details of the laser parameters are given in Appendix~\ref{appendix:1S0}. 
To avoid anti-trapping of the Rydberg state and differential light shifts on the Rydberg transition caused by the tweezer light, we adiabatically ramp to intensity-modulated tweezers with a modulation period of $\SI{1.8}{\mu s}$~\cite{zhang2025leveragingerasureerrorslogical}. During the trap off period, we perform Rydberg excitation followed by an autoionization pulse with a $\SI{369}{nm}$ laser driving a transition to a $6p\nu d$ state. The Rydberg state is autoionized with a $1/e$ removal time of $\tau_{1/e}=\SI{170}{ns}$. To ensure complete removal of Rydberg population, the autoionization pulse remains on until the next Rydberg excitation pulse, with a pulse duration of $\sim 3\mu$s.
During the autoionization time, we also perform optical pumping again to transfer population that decayed to $\ket{^1S_0, m_F = +1/2}$ back to $\ket{g}$. To allow enough time for optical pumping, Rydberg excitation is performed during every other trap-off period. Each cycle consisting of a Rydberg $2\pi$ pulse, autoionization and optical pumping takes a combined $\SI{3.6}{\mu s}$~(Fig.~\ref{fig:fig2}(b)). 

To characterize the protocol performance and probe collective Rydberg dynamics, we measure the atom occupancies in the optical tweezers by performing high-speed number-resolved imaging on the $^1S_0\to{^1P_1}$~($\SI{399}{nm}$) transition in $\SI{10}{\mu s}$~\cite{falconi2025microsecond}. 
The imaging time is much shorter than the timescale of light-assisted collisions by the imaging light, allowing the atom-number distribution in the tweezers to be measured before collisional loss occurs. An example single-shot image of the stochastically loaded $4\times25$ array is shown in Fig.~\ref{fig:fig2}(c), while a photon-count histogram is presented in Fig.~\ref{fig:fig2}(d). 
We extract the atom number distribution by fitting the photon-count histogram with a sum of seven Gaussian functions. The probability $P(N)$ is obtained from the enclosed area of each Gaussian function.

\subsection{Collective intra-tweezer Rydberg dynamics}
An example of single-atom Rabi oscillations is shown in Fig.~\ref{fig:fig2}(e), where the Rabi frequency is $\Omega/2\pi = \SI{4.6}{MHz}$. The observed oscillations agree with a master-equation simulation using experimental parameters, and the visibility of the Rabi oscillations is limited by intermediate state scattering~($T_2^{\mathrm{Rabi}} = \SI{2.7}{\mu s}$) and the Rydberg-state lifetime~(7~$\mu$s). 

To confirm that multiple atoms are blockaded within a tweezer, we measure the atom number distribution as a function of the Rydberg excitation duration~(Fig.~\ref{fig:fig2}(f)). 
The results exhibit complex multi-frequency Rabi oscillations, where the oscillation of $P(N)$ has components of $\sqrt{N}\Omega$ and $\sqrt{N+1}\Omega$.
The complex multi-frequency oscillations are well captured by a master-equation-based simulation that incorporates the single-atom dynamics and Rydberg blockade, with an extra phenomenological dephasing rate $\Gamma_N$ for atom numbers $N\ge 2$~(details of the simulation are discussed in Appendix~\ref{appendix:1S0}). This multi-atom decoherence is also observed in Ref.~\cite{Ebert2014Atomic, Ebert2015Coherence}, where possible explanations include short-range effects from ground-Rydberg molecules or Rydberg \textit{spaghetti} and long-range effects such as finite blockade \cite{Gaj2014molecular,legrand2025revealingelectronytterbiuminteractionsrydberg,Derevianko2015Effects}.

\subsection{Single-atom preparation performance}
Now, we turn to demonstrating single-atom preparation using the approach. Fig.~\ref{fig:fig3}(a) shows the probability to find zero, one or more than one atom in the tweezers after a variable number of repetitions, starting from $\overline{N}=2.3$ atoms and using a detuning of $\Delta/\Omega = 0.75$. After $m=18$ cycles (total duration $\SI{64.8}{\mu s}$), the multi-atom probability reaches $P_{\mathrm{th}} = 1\%$ while maintaining a single-atom filling fraction of $58.2(2)\%$. For comparison, LAC using the same initial conditions and trap depth achieves a similar filling fraction at 1\% multi-atom occupancy ($53.9(2)$\%), but requires $\SI{10}{ms}$~(Fig.~\ref{fig:fig3}(b)). Lowering the multi-atom threshold to 0.1\% requires additional excitation cycles, increasing the protocol duration to $\SI{165.6}{\mu s}$ and reducing the final filling fraction to $35.6(4)\%$. This reduction in the filling fraction arises from single-atom loss, which is independently measured to be $2.0(2)\%$ per cycle for the detuning $\Delta/\Omega=0.75$, reduced from $5.4(1)\%$ per cycle for on-resonant pulses due to decreased intermediate-state scattering and less time spent in the Rydberg state.

To understand the population evolution, we compare the measurements with master-equation simulations~(dashed line in Fig.~\ref{fig:fig3}(a)).
However, we observe two effects not captured by the model. First, as shown in Fig.~\ref{fig:fig3}(a), the measured zero-atom population increases faster than predicted from single-atom loss alone, which can be empirically reproduced by including an $8\%$ probability of $2\rightarrow0$ loss per cycle in the simulations~(solid lines in Fig.~\ref{fig:fig3}(a)). Second, we observe that the removal of multi-atom population slows down later in the protocol~(inset, Fig.~\ref{fig:fig3}(a)), resulting in additional pulses compared to simulations for reaching lower multi-atom thresholds. Neither effect is understood.

The ultimate filling fraction, $P^*(1)$, is limited by the initial filling, single-atom losses, and two-atom losses. We study this systematically by varying the initial loading $\overline{N}$ by changing the tweezer loading time. As shown in Fig.~\ref{fig:fig3}(c), $P^*(1)$ initially increases with $\overline{N}$, reflecting the decreasing initial empty-tweezer probability $e^{-\overline{N}}$, but saturates near $\overline{N}=2$. This trend is reproduced qualitatively by our simulations as above. 
We observe that the experimentally achieved filling fractions are systematically lower than the simulations, which we attribute to the simulations not capturing the observed slowdown described above. In addition, the discrepancy between the circle and star data arises from fluctuations in the 394~nm laser power, which changes the resonance and calibrated $2\pi$ pulse duration, leading to slightly higher single-atom loss for the circle data.

We find that prior cooling of the atoms in tweezers is not required for the protocol, avoiding additional temporal overhead.
The atoms have a temperature of $\SI{80}{\mu K}$, as set by the reservoir temperature. Cooling the atoms to $\SI{20}{\mu K}$ (requiring $\SI{1}{ms}$) before the protocol does not significantly modify the observed dynamics, apart from additional LAC-induced losses during Doppler cooling. 
Although additional cooling is required for subsequent coherent qubit operations, this cooling can be incorporated into the identification imaging step performed after the protocol and therefore does not introduce additional temporal overhead.

\section{Single-atom preparation with single-photon Rydberg excitation}

Further reducing the single-atom loss with two-photon excitation requires both suppressing intermediate-state scattering and choosing a Rydberg state with higher principal quantum number $\nu$, which has a longer lifetime. Using the intermediate state $^1P_1$ in ytterbium, this is challenging because of both the difficulty of generating sufficiently high optical power at \(\SI{395}{nm}\) and the broad linewidth of the $^1P_1$ state ($\Gamma/2\pi = 29.1$~MHz). This contrasts with alkali-atom platforms, where low-loss two-photon Rydberg excitation is readily achievable~\cite{evered2023highfidelity}.
We overcome this limitation by using single-photon excitation from the metastable ${}^3P_0$ state~\cite{ma2023a}. 

To prepare atoms in $\ket{{}^3P_0, m_F=+1/2}$, we perform cooling~($\SI{4}{ms}$) followed by optical pumping~($\sim \SI{1}{ms}$) via a Raman transition to the $5d6s \,^3D_1$ state~\cite{ma2023a}~(Fig.~\ref{fig:fig4}(a)). Accidental light-assisted collisions during cooling lead to a sub-Poissonian atom-number distribution even without applying any Rydberg pulses, yet the majority of tweezers remain multiply-occupied~(Fig.~\ref{fig:fig4}(b)). Atom occupancies are measured after the protocol by depumping the $\ket{{}^3P_0, m_F=+1/2}$ atoms back to the ground state. Additional details of the experimental implementation are given in Appendix~\ref{appendix:3P0}. 

We excite to the Rydberg state $\ket{r^\prime}=\ket{\nu=52.31,L=0,F=1/2,m_F=-1/2}$, chosen for compatibility with high-fidelity two-qubit gates~\cite{liu2026}, from $\ket{{}^3P_0, m_F=+1/2}$ using a single-photon transition driven by a $\SI{302}{nm}$ laser. Representative single-atom Rabi oscillations and coherent multi-atom Rabi oscillations between $\ket{{}^3P_0, m_F=+1/2}$ and $\ket{r^\prime}$ with a Rabi frequency of $\Omega/2\pi = \SI{6}{MHz}$ are shown in Fig.~\ref{fig:fig4}(c,d). The single-atom loss is reduced to $0.26(2)\%$ per $2\pi$ pulse because of the absence of intermediate-state scattering and the longer lifetime of the higher-$\nu$ Rydberg state.

The reduced single-atom loss enables higher single-atom filling fractions than in the two-photon implementation. 
We achieve a filling fraction of $74.8(3)\%$ when reaching a multi-atom probability below $P_{\mathrm{th}} = 1\%$ in $m=11$ cycles, and a filling fraction of $72.4(3)\%$ when $P_{\mathrm{th}}=0.1\%$ with $24$ cycles~(Fig.~\ref{fig:fig4}(e)). 
In contrast to the two-photon implementation, where extending the protocol to lower $P_{\mathrm{th}}$ leads to a significant reduction in $P^*(1)$, the single-photon implementation exhibits only a $\sim 2\%$ decrease in filling fraction between these thresholds because of the low single-atom loss. With this approach, filling fractions above $70\%$ can be achieved with an average number of atoms as low as $\overline{N}=1.5$ at $m=0$~(Fig.~\ref{fig:fig4}(f,g)).

The achieved single-atom filling fraction is no longer limited by single-atom loss, but instead by the observed $N\to N-2$ loss, which is evident in the sharp rise of $P(0)$ probability in the first few rounds of the protocol.
The $N=2\to 0$ loss probability is fitted to be $18\%$ per cycle, over $2\times$ larger than the losses observed with the two-photon implementation. The origin of this two-body loss remains unclear. 
Possible mechanisms include blockade violations via \textit{Rydberg spaghetti}~\cite{Derevianko2015Effects} at short interatomic distances,  or effects from Rydberg atoms interacting with nearby $^3P_0$ atoms in the form of molecules or collisions~\cite{Gaj2014molecular,legrand2025revealingelectronytterbiuminteractionsrydberg,schlagmuller2016collisions,geppert2021collisions}.
The time required for cooling and state-preparation of atoms in $^3P_0$ accounts for the majority of the total protocol duration.
However, these overheads are not fundamental to the protocol and could be shortened in future implementations by, for example, starting with colder atomic samples.

\section{Discussion}
We have demonstrated a fast single-atom preparation protocol in optical tweezers, achieving over two orders of magnitude speedup compared to conventional light-assisted collisions by exploiting intra-tweezer Rydberg blockade.
The essential ingredients for this protocol are low-loss Rydberg excitation and fast removal of Rydberg atoms. Low-loss Rydberg excitation can be readily achieved with tweezer-array experiments using the same ingredients necessary to achieve high-fidelity two-qubit gates. Autoionization is a generic feature of alkaline earth atoms~\cite{madjarov2020high,Cao2024Autoionization} but can be replaced by other Rydberg removal techniques such as anti-trapping or field ionization~\cite{potvliege2006photo,Pillet1983Microwave,Gallagher1977Field}.

This protocol can in principle be extended to deterministic loading, due to the $N\to N-1$ ejection mechanism and the high survival probability of single atoms. Operating in the near-deterministic regime could eliminate the need for taking a non-destructive image and rearrangement, removing the other two processes with major temporal overhead in operating a continuous tweezer array. Two-body losses are the main obstacle to achieving higher filling fractions; their origin is not yet understood, and they are observed to be $8\%$ for our two-photon implementation from the ground state and $18\%$ for our single-photon implementation from the metastable state. 
A single-atom filling fraction of $99\%$ can be reached with single and two-atom losses of $0.1\%$ and $0.05\%$, respectively, starting from $\overline{N}=6$ atoms per tweezer (see Appendix~\ref{appendix:deterministic_loading}). This initial loading is lower than our estimate of $\overline{N}\approx 60$ for reaching the same level of determinism with the \textit{keep-one} approach~\cite{Ebert2014Atomic}.

\section{Acknowledgements}
We acknowledge helpful conversations with Genyue Liu, Guillaume Bornet, Bichen Zhang, Deniz Kurdak, and Mingxuan Xiao. 
This work was supported by the Gordon and Betty Moore Foundation (grant DOI 10.37807/gbmf12253), the Army Research Office (W911NF2410358), the National Science Foundation through the CAREER program (PHY-2047620), DARPA MeasQuIT (HR00112490363), the Office of Naval Research (N00014-23-1-2621), and the NSF Center for Robust Quantum Simulation (OMA-2120757).
J.D.T. is a co-founder and shareholder of Logiqal, Inc.

\appendix

\section{Choice of Rydberg states}
\label{appendix:rydberg_state}
\begin{figure}[tb]
\includegraphics[width=\columnwidth]{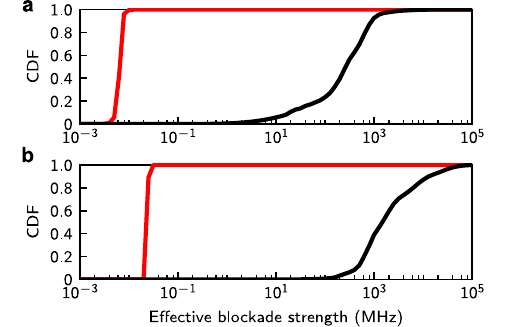}
\caption{\label{fig:figx} Cumulative distribution functions (CDF) of effective Rydberg interaction strengths for atom pairs within the same tweezer (black) and in neighboring tweezers (red): (a) for the $\ket{36.28,L=2,F=5/2}$ state, for a temperature of $T=\SI{80}{\mu K}$ and trap frequencies $(\omega_\mathrm{rad},\omega_\mathrm{ax}) = 2\pi\times(\SI{89}{kHz},\SI{15}{kHz})$, with an inter-tweezer spacing of $\SI{6.25}{\mu m}$; and (b)~for the $\ket{52.31,L=0,F=1/2}$ state for a temperature of $T=\SI{9}{\mu K}$ and trap frequencies $(\omega_\mathrm{rad},\omega_\mathrm{ax}) = 2\pi\times(\SI{65}{kHz},\SI{11}{kHz})$, with an inter-tweezer spacing of $\SI{10}{\mu m}$.}
\end{figure}

We choose the Rydberg states and tweezer spacings such that the interaction strengths are large for atoms within the same tweezer, ensuring strong blockade during Rydberg excitations, while remaining weak for atoms in neighboring tweezers. To estimate the relevant interaction scales under the experimental trapping conditions, we calculate distributions of effective Rydberg interaction strengths 

\begin{equation}
    V_\mathrm{eff}(R,\theta)
    =
    \left(
    \sum_k
    \left|
    \frac{c_k(R,\theta)}{\Delta_k(R,\theta)}
    \right|^2
    \right)^{-1/2},
\end{equation}
where $c_k(R,\theta)=\langle \Phi_k(R,\theta)\mid rr\rangle$ is the overlap amplitude between the interacting pair eigenstate $\ket{\Phi_k(R,\theta)}$ and the target Rydberg pair state $\ket{rr}$, $\Delta_k(R,\theta)$ denotes the corresponding pair-state energy defect, and $\theta$ is the angle between the internuclear axis and the quantization axis defined by the magnetic field. Based on the thermal atomic position distributions of $R$ and $\theta$, the resulting atom-pair-interaction-strength distributions for the two-photon and single-photon excitation schemes are shown in Fig.~\ref{fig:figx}. The spread of the intra-tweezer distributions reflects contributions from both the thermal spread in $R$ and the angular anisotropy of the interaction at 
short interatomic distances.

\section{Details of Rydberg excitation from the ground state}
\label{appendix:1S0}
\subsection{Experimental details }
\label{appendix:experiment_details_1S0}
To excite ground state atoms to the Rydberg state $\ket{r}$, we use a two-photon excitation with typical Rabi frequencies of $\Omega_{399}/2\pi=\SI{130}{MHz}$ (beam power $P=\SI{2}{mW}$, beam waist $w_0=\SI{500}{\mu m}$) and $\Omega_{395}/2\pi=\SI{70}{MHz}$ ($P=\SI{280}{mW}$, $w_0=\SI{50}{\mu m}$), and an intermediate state detuning of $\Delta_e=-\SI{1020}{MHz}$. The sign $\Delta_e\Delta_r<0$ is chosen to mitigate intermediate state scattering \cite{evered2023highfidelity}. 
After Rydberg excitation, an autoionization pulse drives a transition to a $6p\nu d$ state lying energetically above the Yb$^+$ ion core. Approximately $93\%$ of the Rydberg population is ionized ($P_{369}=\SI{18}{mW}$, $w_0=\SI{50}{\mu m}$, $\tau_{1/e}=\SI{170}{ns}$), limited by the finite Rydberg lifetime and scattering of the $\ket{r}$ state back to $\ket{g}$ induced by \SI{395}{nm} photons.
The protocol is performed in 488~nm tweezers with depth of 380~$\mu$K.
The magnetic field of \SI{8.5}{G} is oriented perpendicular to the tweezer axial direction.
Due to the orientation of the beams with respect to the $4\times25$ tweezer array, we only select out 50 traps lying along the direction of propagation of the beams for analysis of experiments involving Rydberg dynamics.  To evaluate the protocol performance when multi-atom probabilities are less than a few percent, we employ thresholding rather than fitting $P(N)$ by Gaussian functions.

\subsection{Simulation details}
\label{appendix:1s0_simulation}
For the two-photon Rydberg excitations from $^1S_0$, we model the Rabi oscillations and protocol with master-equation-based simulations with collapse operators to account for intermediate state scattering and finite Rydberg lifetime. For multiple atoms within a tweezer, we account for the Rydberg blockade with a $V/2\pi=\SI{100}{MHz}$ interaction energy for pairs of atoms in the Rydberg state.

Our measurements of the $^1S_0$ population also include Rydberg atoms that are not properly autoionized, and return to the ground state via decay or scattering.
We account for this in simulations by multiplying the Rydberg population at the end of the Rydberg pulse by a fixed probability of autoionizing a Rydberg atom. The autoionization probability is $93\%$, which was independently computed based on measurements of the autoionization timescale, Rydberg lifetime, and inferred $\SI{395}{nm}$ scattering. This probability is used for Fig.~\ref{fig:fig2}(e) and Fig.~\ref{fig:fig3}, however, we find a better fit to the data in Fig.~\ref{fig:fig2}(f) with an autoionization probability of $90\%$, likely as a result of fluctuating $\SI{369}{nm}$ laser intensity.

For multiple atoms within a tweezer, we include an extra dephasing collapse operator $\sqrt{\Gamma_N}\sigma_z$. We fit $\Gamma_N$ for $N=2$ through $N=4$ to minimize the residuals between the simulated values and experimental data from Fig.~\ref{fig:fig2}(f). The fitted values are $\Gamma_2/2\pi=\SI{0.116(3)}{MHz}$, $\Gamma_3/2\pi=\SI{0.194(6)}{MHz}$, and $\Gamma_4=\SI{0.18(3)}{MHz}$.

\section{Details of Rydberg excitation from the $^3P_0$ state}
\label{appendix:3P0}
\subsection{Experimental details}
\label{appendix:experiment_details_3P0}

In order to limit atom losses during the preparation and readout of $^3P_0$, it is necessary to cool the atoms using gray-molasses cooling detailed in Ref.~\cite{li2025fast}. The atomic temperature of the $^3P_0$ atoms when the protocol is performed is 9~$\mu$K.
Light-assisted collisions during the cooling process result in a sub-Poissonian distribution of the atom number, as well as higher temperatures for multiple atoms in a trap due to collisional heating. 
The sub-Poissonian atom distribution after cooling, pumping to $^3P_0$, and depumping can be modeled by a Mandel parameter $Q=\frac{\sigma_{\overline{N}}^2}{\overline{N}}-1$ ranging from $0$ to $-0.25$ for $\overline{N}=0$ through $\overline{N}=2.0$. For simulations in Fig.~\ref{fig:fig4}(f,g), we interpolate $Q$ linearly in this range to approximate the initial atom distributions prior to Rydberg evolution.

After cooling, atoms are prepared in the metastable state $\ket{^3P_0, m_F = +1/2}$ using optical pumping. First, we apply a global $\sigma^{+}$ 556~nm beam that drives the $\ket{^1S_0,m_F=-1/2}\rightarrow\ket{^3P_1,m_F=+1/2}$ transition to prepare the atoms in $\ket{^1S_0,m_F=+1/2}$ state. Then, we coherently excite atoms to $5d6s\,^3D_1$ using a two-photon Raman transition $\ket{^1S_0,m_F=1/2}\rightarrow\ket{^3P_1,m_F=3/2}$ (556~nm) and $\ket{^3P_1,m_F=3/2}\rightarrow \ket{5d6s\,^3D_1,m_F=3/2}$ (1539~nm). 
The $5d6s\,^3D_1$ state decays into the desired $\ket{^3P_0, m_F = +1/2}$ state, as well as $6s6p\,^3P_1$ and $6s6p\,^3P_2$. $6s6p\,^3P_1$ decays back to $^1S_0$ and is pumped again, and $6s6p\,^3P_2$ is constantly depumped through the $6s6d\,^3D_2$ state using a 497~nm laser. 
The $^3P_0$ atoms are trapped in 488~nm tweezers with trap depth of 220~$\mu$K. The magnetic field of \SI{75}{G} is oriented perpendicular to the tweezer axial direction.
To readout the atom population, we use the spin-selective depumping scheme detailed in Ref.~\cite{li2025fast} to only transfer atoms in $\ket{^3P_0, m_F = +1/2}$ back to $^1S_0$ prior to number-resolved imaging.

A 302~nm laser with $\sigma^-$ polarization, power of $P=\SI{73}{mW}$ and beam waist of $w_0=\SI{20}{\mu m}$ is used to drive the $\ket{^3P_0, m_F = +1/2}$ to the Rydberg state $\ket{r^\prime}$.
Since this $\ket{r^\prime}$ state has a higher $\nu$ compared to the state chosen for the ground state excitation, the interaction strength and blockade radius are larger, so we increase the tweezer spacing to $\SI{10}{\mu m}$. As a result of this larger spacing and the smaller beam waist of the $\SI{302}{nm}$ laser, the protocol performance is characterized in a 1D array of 7 atoms. The autoionization removal time of the $\ket{r'}$ state is $\tau_{1/e}=\SI{44(5)}{ns}$.

\subsection{Simulation details}
\label{appendix:3p0_simulation}

For the single-photon Rydberg excitations from $^3P_0$, in both the multi-atom Rabi oscillations and single-atom preparation data we observe that the $N=0$ population increases by more than expected from just the single-atom losses of $0.26(2)\%$. The population evolutions are consistent with an additional $N\to N-2$ loss channel. Thus, we include a two-atom loss modeled with an exponential decay, with phenomenological rate $\tau_{2b}^{(N)}$, which we fit for each $N\ge 2$.
From the measured autoionization timescale of $\SI{44(5)}{ns}$ and Rydberg lifetime of $\SI{43(2)}{\mu s}$, we estimate a $99.9\%$ probability of autoionizing Rydberg atoms.

\section{Deterministic loading}
\label{appendix:deterministic_loading}

\begin{figure}[tb]
\includegraphics[width=\columnwidth]{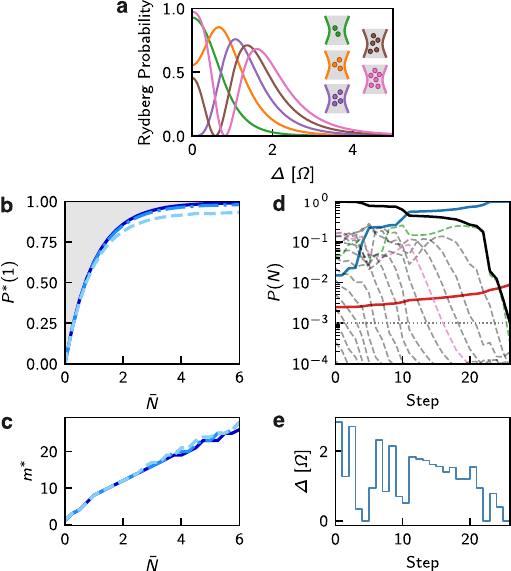}
\caption{\label{fig:fig5} 
(a)~The probability of exciting a single atom to the Rydberg state for $N$ blockaded atoms as a function of detuning $\Delta$ at the single-atom $2\pi$ pulse time $T=2\pi/{\sqrt{\Omega^2+\Delta^2}}$, for $N=2$ through $N=6$. 
(b)~The single-atom probability, $P^*(1)$, after reaching $P(>1)<0.1\%$ as a function of $\overline{N}$, for sequences of optimized detunings. The three curves correspond to different single ($L_1$) and two-atom ($L_2$) loss probabilities of $\{L_1=0.1\%,L_2=0.05\%\}$ (solid dark blue), $\{L_1=0.25\%,L_2=0.25\%\}$ (dot-dashed medium blue), and $\{L_1=1\%,L_2=1\%\}$ (dashed light blue). 
(c)~The number of cycles required ($m^*$) to reach $P(>1)<0.1\%$ as a function of $\overline{N}$, for the sequences corresponding to the curves in (b).
(d)~The probability evolution for reaching a single-atom filling fraction of $99\%$ with $\overline{N}=6$ with a single-atom loss probability of $0.1\%$ and two-atom loss probability of $0.05\%$ (dark blue curve in (b-c)). The solid red, blue, and black lines correspond to $N=0$, $N=1$, and $N>1$. The dashed curves depict $P(N)$ values for individual $N>1$, with the green and pink curves for $N=2$ and $N=6$, respectively, highlighted. 
(e)~The sequence of detunings used at each step to produce the population evolution in (d).
}
\end{figure}

In our experimental results, we do not observe population accumulation at certain atom numbers $N$ (i.e. $N=4,9,...$ for $\Delta=0$), despite operating at a single static detuning. This is a result of ground-Rydberg multi-atom dephasing and two-atom losses. Should these multi-atom Rydberg effects be mitigated in future experiments, the question of optimal pulse sequencing becomes relevant for quick multi-atom removal that prevents population accumulation at certain $N$. For deterministic loading in the presence of small single and two-atom losses, it is important to control the population evolution so that a majority of the population is successively moved from higher to lower $N$. One mechanism for controlling the population evolution is to use different detunings at each cycle, in order to engineer higher (or lower) $N\to N-1$ transition rates (see Fig.~\ref{fig:fig5}(a)). 

Therefore, we simulate protocols with different detunings for each pulse, based on analytic expressions for superatom Rabi oscillations with small single and two-atom losses. The single-atom losses are modeled based on the Rydberg lifetime, and two-atom losses are modeled as proportional to the pulse length.
Given some initial mean number of atoms loaded, $\overline{N}$, and single and two-atom loss probabilities, we find an optimal sequence of detunings $S=\{\Delta_1,...,\Delta_{m}\}$ by minimizing a cost function
\begin{equation}
C(S)=-P(1)_S+k_1m+k_2\theta(P(>1)_S-P_{\mathrm{th}}),
\end{equation}
where $P(1)_S$ and $P(>1)_S$ are the single and multi-atom probabilities after $m$ pulses, $P_{\mathrm{th}}$ is a multi-atom threshold, and $\theta$ is the Heaviside step function. The constants $k_1=0.001$ and $k_2=10^3$ penalize longer sequences and sequences that do not yield a multi-atom probability below the threshold, respectively. The first term $-P(1)_S$ maximizes the final single-atom probability. Because $k_1=0.001$, a sequence that requires an extra pulse compared to another sequence is only deemed more optimal if it increases $P(1)$ by more than $0.1\%$.

In Fig.~\ref{fig:fig5}(b-c), we present $P^*(1)$ and $m^*$ as a function of $\overline{N}$, for different single and two-atom loss probabilities. These loss probabilities are given for on-resonant pulses.
A single-atom filling fraction of $99\%$ can be reached with a single-atom loss probability of $L_1=0.1\%$, two-atom loss probability of $L_2=0.05\%$, $\overline{N}=6$, and $m=26$. The population evolution and detuning sequence for reaching $99\%$ are shown in Fig.~\ref{fig:fig5}(d-e). Meanwhile, $99.9\%$ can be reached with $L_1=0.008\%,L_2=0.005\%,\overline{N}=10$, and $m=39$.

\bibliography{bib}

\end{document}